# Towards a direct measurement of the quantum-vacuum Lagrangian coupling coefficients using two counterpropagating super-intense laser pulses

Luis Roso[1,2], Roberto Lera[2], Smrithan Ravichandran[3,4], Andrew Longman[5*], Calvin Z He[3,4], José Antonio Pérez-Hernández[1], Jon I Apiñaniz[1], Lucas D Smith[6], Robert Fedosejevs[5], and Wendell T Hill, III[3,4,6]

1 Centro de Láseres Pulsados (CLPU), 37185 Villamayor, Salamanca, Spain
2 Departamento de Física Aplicada, Universidad de Salamanca, 37008 Salamanca, Spain
3 Joint Quantum Institute, University of Maryland, College Park, MD 20742, USA
4 Institute for Physical Science and Technology, University of Maryland, College Park, MD 20742, USA
5 Electrical and Computer Engineering, University of Alberta, Edmonton, Alberta T6G 2V4, Canada
6 Department of Physics, University of Maryland, College Park, MD 20742, USA
* Currently at Lawrence Livermore National Laboratory, Livermore, CA 94550, USA

Email: wth@umd.edu

**Abstract**

In this paper we will show that photon-photon collision experiments using extreme lasers can provide measurable effects giving fundamental information about the essence of QED, its Lagrangian. A possible scenario with two counterpropagating ultra-intense lasers for an experiment to detect scattering between optical photons is analyzed. We discuss the importance of the pulse widths and waists, the best scenario for overlapping the beams and signal detection, as well as ways to distinguish the signal from the noise. This would need a high-precision measurement, with control of temporal jitter and noise. We conclude that such experiment is barely feasible at $10^{23}$ W/cm$^2$ and very promising at $10^{24}$ W/cm$^2$.

## 1.- Introduction.

Concomitant with early efforts to unify quantum mechanics, special relativity and electrodynamics, were suggestions of creating electron-positron pairs during the collision of two suitably energetic photons [1], $\gamma\gamma \rightarrow e^+e^-$. The observation of this process, often referred to as Breit-Wheeler pair production [2], has proven to be illusive due to the incredibly small cross section even under the best conditions, $\hbar\omega_\gamma \sim 0.5$ MeV $\rightarrow \sigma_{\gamma\gamma} \sim 10^{-30}$ cm$^2$. By contrast, the reverse process, electron-positron annihilation also suggested by Dirac, is readily observed in high-energy physics experiments. Since the development of the laser, contemplation has largely focused on a variety of forms of the so-called multiphoton Breit-Wheeler pair production. The one and only example to date is the SLAC E-144 experiment [3], which exploited inverse Compton scattering to frequency-shift infrared laser photons, $\omega_L$, into gamma photons, $\omega_\gamma$, leading to $\hbar\omega_\gamma + n\hbar\omega_L \rightarrow e^+e^-$. This experiment was performed in the perturbative regime where the invariant nonlinearity quantum parameter, $\chi$, is below threshold, i.e.,

$$\chi = \frac{2\hbar\omega_\gamma}{m_e c^2} \frac{E}{E_{cr}} < 1 \qquad (1)$$

where $m_e$ indicates the electron mass and $E$ and $E_{cr}$ are the effective peak electric field at focus and the critical field, $m_e^2 c^3 / e\hbar \sim 1.3 \times 10^{20}$ V/m [4, 5, 6, 7] required to separate $e^+e^-$ pairs by the Compton wavelength, respectively. The small number of positrons generated was clear evidence of the process. Today, several groups are planning updated versions of this experiment with $\chi > 1$ that are expected to lead to copious numbers of pairs.

Besides electromagnetic waves, it is possible to just consider electrostatic fields, as in the initial proposal of Schwinger [7] or magnetic fields [8, 9, 10, 11], without laser fields, but in these static cases laboratory experiments seem unlikely in the next years.

There is evidence of Delbrück scattering of high energy photons in the Coulomb field of a nucleus [12, 13, 14, 15]. More recently direct $\gamma\gamma \to \gamma\gamma$ has been observed via ultraperipheral collisions between Pb nuclei at the Large Hadron Collider [16].

Despite the fact that there is little doubt of the reality of the Breit-Wheeler mechanism, there is a paucity of direct experimental evidence of what happens before positrons are coaxed out of the negative energy sea, i.e., when $\chi \ll 1$ but the other invariant parameter (sometimes referred to as the classical nonlinearity parameter), $a_0$ is large, i.e.,

$$a_0 = \frac{eE}{2\pi m_e c \omega_L} \gg 1 \qquad (2)$$

In this regime, photons can interact with each other, and classical electrodynamics (Maxwell equations) no longer adequately describe the physics. Theories beyond Maxwell, incorporating quantum mechanics and the nonlinear interaction between photons, are required to describe the physics in this sector. We are interested in the so-called box diagram describing the elastic scattering of two photons, $\gamma\gamma \to \gamma\gamma$, with electron-positron pairs as intermediaries. The ultraperipheral collisions observed at the Large Hadron Collider [16] found a cross section in agreement with traditional model predictions. This is very relevant from a fundamental point of view for a variety of reasons. First and foremost is a need to understand the nature of the quantum vacuum (qvac) and the correct Lagrangian to describe it. At the same time, the hypothetical existence of new particles, e.g., axion-like particles, mini-charged particles with masses < $m_e$, would influence the qvac behavior long before the onset of ordinary pair production. Quantitative knowledge of such particles and how they affect the qvac could shed light and place limits on the types of additions that can be made to The Standard Model of Particle Physics.

Most often, contemporary literature discusses the qvac and its Lagrangian in terms of quantum electrodynamics (QED), but there are other formulations. Prior to the development of the lowest order QED expression, which is identical to the Heisenberg-Euler model [5], for example, Born-Infeld [17] presented a Lagrangian to address the infinite self-energy of the electron. As we will see below, the two models are similar in appearance but differ drastically in what they predict about the qvac. Of particular note is that the leading nonlinear terms of QED suggest a 7:4 qvac birefringence in the phase in the presence of a strong laser field (that translated to scattered intensity results in a 49:16 signal ratio) whereas those of the Born-Infeld formulation lead to an isotropic qvac. In addition, the magnitude of the QED birefringence depends on the reality or absence of the hypothetical particles mentioned above.

Without direct experimental evidence in the strong-field regime (i.e., $a_0 \gg 1$ but $\chi \ll 1$) it is hard to state with confidence which, if any, model is consistent with reality[1]. To enhance the signal we are looking for, we need to reach the high intensity regime. Even though there is a plethora of indirect support for QED in the weak-field regime, because the photon-photon interactions are so weak and the coupling constants are so small (see below), it will take a precision measurement in the strong-field regime to distinguish the different models.

While there is no conclusive experimental evidence of the birefringence of the qvac, there is indirect evidence [18]. Theoretically, birefringence should also occur with super-strong magnetic fields below a critical value of $4.4 \times 10^{13}$ G [8, 9, 10]. Perhaps the most sensitive and precision lab-based polarimetry measurements have been in the optical region by the PVLAS collaboration [11]. These experiments look for tiny polarization rotation due to the birefringence. While, sensitivity continues to improve, it is not yet at the level to confirm or rule out the birefringence QED predicts. Because the cross section drops precipitously the more the photon energy differs from the rest energy of the electron, there are also efforts underway to measure the birefringence with X-ray probes, where the cross sections are higher, in the presence of strong laser fields [19, 20]. In the quest for simplicity there is a very interesting experimental scenario presented by Paulus and co-workers [21] where they study a self-crossing taking profit of the high repetition rate of the EuXFEL (European X-Ray Free Electron Laser). Experiments are also in the planning stage at the Shanghai Coherent Light Facility [22] as well as at the Japanese SACLA [23, 24].

Polarimetry measurement depends on the qvac having a birefringence – a difference in the parallel and perpendicular indices of refraction. Thus, in the absence of a qvac birefringence, polarimetry measurements would not provide any information about the magnitude of the coupling constants of the Lagrangian. This leaves room for investigating the qvac in ways that do not require qvac to be birefringent. We point out, that if axion-like or mini-charged particles do exist, some version of them could modify QED to reduce the birefringence of the qvac significantly [25], again pointing out the need for alternative ways to investigate the Lagrangian and the need for precision measurements. To that end, several schemes relying on the interactions between photons mediated by the vacuum have been proposed. These range from direct scattering between x-rays [26] to 4-wave mixing in the near infrared [27] to beam deflection in the visible [28]. Thus far, none of these have had sufficient sensitivity to see the effect predicted by QED. However, like the PVLAS experiments, they have set upper limits to the cross section.

---

[1] We point out that $a_0 \gg 1$ is only required for a certain class of experiments to which the experiment in this manuscript belongs. There are several experiments proposed and under way that do not require the strong-field condition to be met. In the strong-field regime, direct observation of scattered photons as proposed is possible.

Tommasini and coworkers proposed a simple experiment to measure the qvac Lagrangian nonlinear parameters via photon-photon scattering with two counterpropagating laser beams in the near infrared [29, 30, 31]. It is interesting to note that a scattering experiment was first suggested by Halpern [32]. Tommasini's analysis showed the approach to be promising, but the analysis was based on ideal conditions, with no accommodations for experimental factors. In the last decade lasers have evolved a lot, and there are a number of petawatt (PW) laser facilities in operations around the world, some of them with a high repetition rate, ~ 1 shot/s [33], and a few multi-PW (as high as 10 PW today) facilities either being commissioned or under construction. Moreover, due to the need for pump probe experiments in many applications, most of the multi-PW lasers have a PW-class synchronized partner. So, it is time to revisit the idea and to propose a design for performing conclusive experiments at one of these facilities, which is the purpose of this manuscript. Photon-photon scattering experiments are distinct from polarimetry experiments in that they do not rely on the qvac being birefringent in order to measure a signal.

We will present a scenario employing two big counterpropagating lasers colliding. This experimental set-up is arguably the simplest one to show photon-photon scattering nonlinearities [34, 35, 36]. Nevertheless, it presents several technical challenges. Other schemes have been proposed involving overlapping of three or four beams [37]; the effect of such combined nonlinearities is very appealing but the experimental complications for precise synchronization of multiple beams grow exponentially. For this reason, we focus on a simpler scenario.

## 2.- Effective Lagrangians for photon-photon coupling in the optical region.

It is widely accepted that the QED Lagrangian for the low frequency electromagnetic fields ($\hbar\omega \ll m_e c^2$) to first-order correction is given by [38]:

$$\mathcal{L} = \mathcal{L}_0 + \xi_L \mathcal{L}_0^2 + \xi_T (7/4)\mathcal{G}^2 \quad (3)$$

where,

$$\mathcal{L}_0 = \varepsilon_0 (E^2 - c^2 B^2)/2 \text{ and } \mathcal{G} = \varepsilon_0 c(E \cdot B) \quad (4)$$

represent the only covariant combinations of E and B fields possible. Although this Lagrangian applies for static fields, it is acceptable for slowly varying fields as the optical frequencies considered in this paper [39]. Because the scattering cross section is expected to be extremely small, it is appropriate to consider just the lowest order nonlinear terms in our analysis.

The key point for understanding the motivation of this paper is that the nonlinear terms are weighted by these two coefficients $\xi_L$ and $\xi_T$. In their theory Heisenberg and Euler [5] showed the instability of the vacuum for high fields based on this Lagrangian with the coupling constants

$$\xi_L^{QED} = \xi_T^{QED} = \xi^{QED} = \frac{8\alpha^2 \hbar^3}{45 m_e c^5} = 6.7 \ 10^{-24} \text{ cm}^3/\text{J} \quad (5)$$

The qvac Lagrangian (1) with these coupling constants is considered the basis of present-day QED. This was reinforced by the elegant work of Schwinger [7]. However, as mentioned above there are other options. The Born-Infeld formulation [17], for example, predicts a different relationship between the two coupling coefficients, $4\xi_L^{BI} = 7\xi_T^{BI}$. That is, the Born-Infeld theory predicts a line in the $(\xi_L, \xi_T)$ plane, while Heisenberg-Euler (QED) not only predicts a qvac birefringence, it provides a specific magnitude for the coupling constants. The two formulations give incompatible results for the coupling terms: observe that the point $(\xi_L^{QED}, \xi_T^{QED})$ does not lie on the Born-Infeld line, see [40], for example. Clearly, Born-Infeld and Heisenberg-Euler (and QED) are in conflict, albeit the latter has prevalence.

There are other possible coupling terms, to this order, due to the other conjectured contributions such as axions [11], or perhaps dark matter [41]. The discrepancy between the Heisenberg-Euler Lagrangian, and other Lagrangians, as the Born-Infeld, is a basic fundamental problem that has not been properly considered for many decades. Better knowledge could be key to better understanding of some of the present-day questions that The Standard Model of Particle Physics is not able to explain. A direct optical measurement of these coefficients would be of great interest and is the primary objective of the study described in this paper.

Because we are going to focus on these two coupling coefficients, we prefer to write the basic Lagrangian in the following form:

$$\mathcal{L} = \mathcal{L}_0 + q_L \xi_L^{QED} \mathcal{L}_0^2 + q_T \xi_T^{QED} (7/4)\mathcal{G}^2 \quad (6)$$

where, for simplicity, we introduced two dimensionless parameters $q_L$ and $q_T$ that account for the possible deviation from the Heisenberg-Euler Lagrangian (i.e., QED). Obviously,

$$q_L = \xi_L / \xi_L^{QED} \text{ and } q_T = \xi_T / \xi_T^{QED} \quad (7)$$

In this paper we analyze how to perform an experiment to measure the coefficients $q_L$ and $q_T$ directly from the interaction of two linearly polarized ultra-intense lasers: $q_L$ for parallel and $q_T$ for perpendicular alignment of the polarizations of the two lasers.

Tommasini and coworkers [29, 30, 31] showed that Halpern's idea of two counterpropagating lasers can be a good scheme to measure the Lagrangian nonlinear coefficients, $q_L$ and $q_T$. Their main result is that a pump beam will induce a phase change on the wavefront of a probe beam and thus the probe is diffracted. (Alternatively, one can think of the strong laser as polarizing the vacuum virtual pairs as one would a dielectric in ordinary optics). The phase change is directly linked (as they showed) to the QED coupling parameters so the qvac effect can be experimentally seen as proportional to the phase change, and therefore generates a characteristic diffraction pattern. Of course, the effect is very weak, and the experiment must be designed very carefully. In their 2014 paper [25] an experiment was outlined for Gaussian pump and probe beams. However, for the most intense possible pump beam, a high-quality Gaussian profile may be too demanding.

In the present paper we review the Tommasini model and adapt it to a more feasible experimental situation based on existing and planned multi-petawatt lasers [33]. We will present what we consider the best experimental setup within the realistic expectations of world class multi-petawatt lasers. Noise sources are very relevant for such a delicate experiment and are also going to be described. In this paper we will make realistic estimates of the count rates, in presence of background and noise, for lasers expected to be in full operation in the near term.

**3.- Geometry of a possible experiment.**

We will limit our discussion to exact counter-propagating beams, although a small angle would help in terms of the safety for the laser components and would not invalidate our results.

Consider a pump pulse A and a probe pulse B. Beam A must be linearly polarized and moderately tight focused to enhance the nonlinear effects. A very tight focus would mix perpendicular and parallel polarizations and thus will not allow the observation of the effect we are interested in. We will restrict our investigation in this paper to a moderately tight focus to keep the paraxial approximation valid.

While the shape of A is not important, that of B must be as close to Gaussian as possible, with rapidly falling wings away from the axis. Beam B should have a relatively wide beam waist and a very good $TEM_{00}$ structure to keep its linear diffraction wings at a quite low level. Its polarization must be linear too, with the possibility of being parallel or perpendicular to the pump polarization.

Both focal spots must coincide in space and time. Because we consider a wide probe waist, spatial superposition is easy to achieve. The Rayleigh length of the probe will therefore be relatively long. However, the Rayleigh length (or the focus length) of the pump is going to be small. We are interested in the regime where the effective pulse lengths of both beams in the focus are smaller than the pump Rayleigh range. Temporal overlapping must be complete and for that all pump energy has to be close to its waist at the same time. This condition requires femtosecond (~20 to 30 fs) pulses to be employed both for pump and probe, and Rayleigh lengths of the pump of the order of 20 wavelengths or more.

One advantage of relaxing the focus is that the polarization components of the field remain distinct allowing their contributions to be isolated and measured independently. Within paraxial approximation it is well known [42, 43, 44] that there is a longitudinal component of the field at the waist (to satisfy the div $\vec{E}=0$ condition). The leading term of this longitudinal field is of the order $1/kw_o$ times the leading transverse component, k being the wavenumber and $w_o$ the waist. In all calculations to be presented within this manuscript we will keep $1/kw_o<0.1$, both for the pump and the probe, and thus the influence of the longitudinal component can be neglected. This is consistent with the paraxial approximation, when $1/kw_o$ is not small the paraxial approximation completely fails.

This experiment is a targetless experiment in the sense that we don't need to place anything in the beam. Just the opposite, even the residual chamber gas can be a problem (as we will discuss below). We will discuss how to mitigate its contribution in Sec. 7.

*3.1 Counterpropagating fields.*

We will define z along the longitudinal (propagation) direction of beam B, and x, y directions in the transverse plane. In all cases linearly polarized lasers will be considered. The fields, while interacting close to the focal plane, are of the form

$$\tilde{E}_A(x,y,z,t) = E_A(\rho) \cos(k(z-ct)) f(z-ct)$$
$$\tilde{E}_B(x,y,z,t) = E_B(\rho) \cos(k(z+ct)) f(z+ct) \quad (8)$$

where f(z) is the temporal profile of the pulses (normalized to f=1 at the peak), which we assume are the same. Both A and B are linearly polarized but their respective polarizations can be parallel or perpendicular, to explore the vacuum birefringence. For simplicity both beams have cylindrical symmetry ($\rho^2 = x^2 + y^2$), that is reasonable in most cases, and femtosecond duration (typically about 20 to 30 fs).

The electric field amplitudes ($E_A$, and $E_B$) can be complex and may represent the wavefront structure of those fields but are independent of z. This is reasonable only very close to the focus or the waist of the two beams (at z=0).

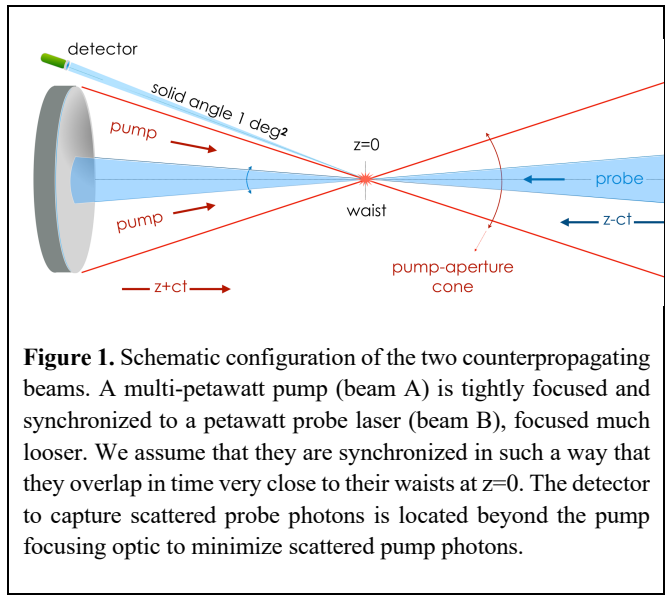

**Figure 1.** Schematic configuration of the two counterpropagating beams. A multi-petawatt pump (beam A) is tightly focused and synchronized to a petawatt probe laser (beam B), focused much looser. We assume that they are synchronized in such a way that they overlap in time very close to their waists at z=0. The detector to capture scattered probe photons is located beyond the pump focusing optic to minimize scattered pump photons.

The pump-probe configuration is shown schematically in Fig. 1. We also consider a detector able to measure the probe photons scattered by the qvac nonlinearity. For simplicity throughout this paper, we will consider that the detector has an aperture of 1 deg$^2$; scaling to other detector apertures is straightforward.

In spite of being a targetless experiment, the architecture is quite complex because it requires that the two beams encounter each other exactly at the focal plane, i.e., the region of the maximum intensity. Even though the probe will have a wide waist and a relatively long Rayleigh length ($z_B$), the pump must be tightly focused and have a short Rayleigh length. Therefore, the synchronization requirements are very tight. The influence of the pulse temporal delay will be analyzed in section 4. To first order, the mismatch in the waists of the beams will help to mitigate the effects of small beam pointing instabilities.

*3.2 Point to point interaction.*

We consider that both pump and probe fields collide close to the waist (quite realistic for pulses shorter than the Rayleigh length of the focus). In that case we can assume that both wavefronts are flat. Considering a volume element dS dz, with dS being a surface element (in the xy plane, i.e., along the wavefront) and dz a length element along the propagation direction, it is possible to get a simple expression for the differential phase shift of the probe. If the energy density in this volume due to the pump beam is $\rho_e(x,y,z,t)$, then according to Tommasini and coworkers [30] the nonlinear phase change of the probe when going through this volume element is

$$d\phi_L(x,y) = 4\xi_L k_B \rho_e \, dz$$
$$d\phi_T(x,y) = 7\xi_T k_B \rho_e \, dz \quad (9)$$

Observe that $\rho_e dSdz/\hbar\omega_A$ is the number of pump photons inside that volume element ($\omega_A$ being the pump laser central frequency). Integrating over z and taking into account that $\rho_e c$ is the intensity at this volume element, one gets

$$\phi_L(x,y) = 4\xi_L k_B I_A(x,y)\tau_A = 4q_L\xi_L^{QED} k_B I_A(x,y)\tau_A \quad (10)$$
$$\phi_T(x,y) = 7\xi_T k_B I_A(x,y)\tau_A = 7q_T\xi_T^{QED} k_B I_A(x,y)\tau_A$$

where $I_A(x,y)$ is the intensity corresponding to the amplitude $E_A(x,y)$, and $\tau_A$ is the pulse duration. Here we considered a square envelope for the time profile. Other time profiles would give a similar expression with a correction factor of the order of one. Tommasini et al. showed that the coupling constants $q_L\xi_L$ and $q_T\xi_T$ are directly related to a phase change, which will modify the diffraction pattern very slightly. Therefore, from the difference between the probe beam diffraction pattern with and without the pump beam, we can measure experimentally the $q_L$ and $q_T$ coefficients.

It is important to notice that relation (10) holds for each longitudinal pencil of the collision we describe. To calculate the full interaction, we then integrate over the transverse profile of both beams as we present below.

*3.3 Physical interpretation.*

Observe that $I_A(x,y)\tau_A$ is the energy of the pump beam per unit area. In a head-on collision, the phase shift is proportional to the number of photons per unit area (fluence). Because of that, to have an efficient qvac experiment it is necessary that both pulse durations, pump and probe, must be short enough to be entirely in the focal region at the same time. In other words, $c\tau_B \sim c\tau_A < z_A$, $z_A$ being the Rayleigh length (or the depth of focus) of the pump. This condition is met for $w_A$=2 µm when pulses are 30 fs or less. Reducing the pulse duration below this condition, $c\tau_B \sim c\tau_A \ll z_A$, does not represent any benefit and introduces extra complications. We will consider probe waists much bigger and durations of the same order as the pump, so the limiting condition is related to the pump beam waist. Observe that we refer to the pump Rayleigh length while the pump may have an arbitrary profile. Rayleigh length here has to be considered just as a measure of the depth of focus. Therefore, the pulse duration most suitable for such a qvac experiment is when pump and probe both have approximately the same length and when that length is smaller (not necessarily much smaller) than the depth of focus.

## 4.- Numerical model.

The photon-photon interaction is going to happen at the focal plane, or very close to it, but the detection is going to be at a large distance to see the diffraction effects due to the qvac phase shift. Because the distance to the detector will be very large, it is reasonable to assume that the field distribution at the detection plane ($z \to -\infty$) will be the Fourier transform (Fraunhofer diffraction) of the field at the focal plane (z=0).

To understand the effect of the qvac coupling we need to develop first the diffraction integral for the unperturbed probe beam, the B beam, and compare it with the same probe beam but perturbed by the pump laser. The difference will be the qvac signal we are looking for.

### 4.1 Unperturbed probe beam.

The incoming probe beam B is described by its amplitude $E_B(x,y)$ at the z=0 plane. Without the pump beam, and assuming cylindrical symmetry in the transverse profile, the Fraunhofer diffraction pattern of the probe beam will be

$$F_B(\zeta) = -\frac{i}{\lambda_B}\int_0^R 2\pi\rho E_B(\rho) J_0(k_B\rho\zeta) d\rho \quad (11)$$

where $F_B(\zeta)$ indicates the electric field amplitude as a function of the polar diffraction angle, $\zeta$, at the detection plane; $J_0$ is the 0$^{th}$ order Bessel function that accounts for the angular integral. We call it F instead of E for clarity – F is at the detector plane and E is in the focal plane. This expression is simply the 2D Fourier transform (with cylindrical symmetry) of the amplitude in the focal plane [45]. Instead of having the Fourier transform from zero to infinity we stop the integral at radius, R = 4 mm (five thousand wavelengths), which shows that an aperture of this size does not have any effect on the qvac-scattering signal. Continuation of the Fourier integral to larger radii does not change the result.

### 4.2 Intensity at detector with pump beam on.

Now we include the pump beam A synchronized to arrive to the z=0 plane at the same time as the probe. With the A beam on and assuming cylindrical symmetry, the B beam $E_B$ is modified to the $E_S$ due to the phase change.

The schematic representation of the interaction close to the focal plane is shown in Fig. 2. Before the interaction with the qvac perturbed by the pump beam A, the probe beam has a profile given by $E_B(x,y)$. Just after the interaction the wavefront has an added phase $E_S(x,y) = E_B(x,y)\,\phi(x,y)$, one phase for each polarization situation (parallel or crossed). The difference between the diffraction pattern of $E_S$ and the unperturbed $E_B$ diffraction is the evidence for the qvac effect that we want to measure.

The amplitude at the detection plane is given by the polar angle, again due to the cylindrical symmetry. Considering the same aperture of radius R at the focal plane, as in the previous case, one gets:

$$F_S^P(\zeta) = -\frac{i}{\lambda_B}\int_0^R 2\pi\rho E_S^P(\rho) J_0(k_B\rho\zeta) d\rho \quad (12)$$

where the index P indicates the L or T relative polarization of the two beams, as in Eq. (9). Then,

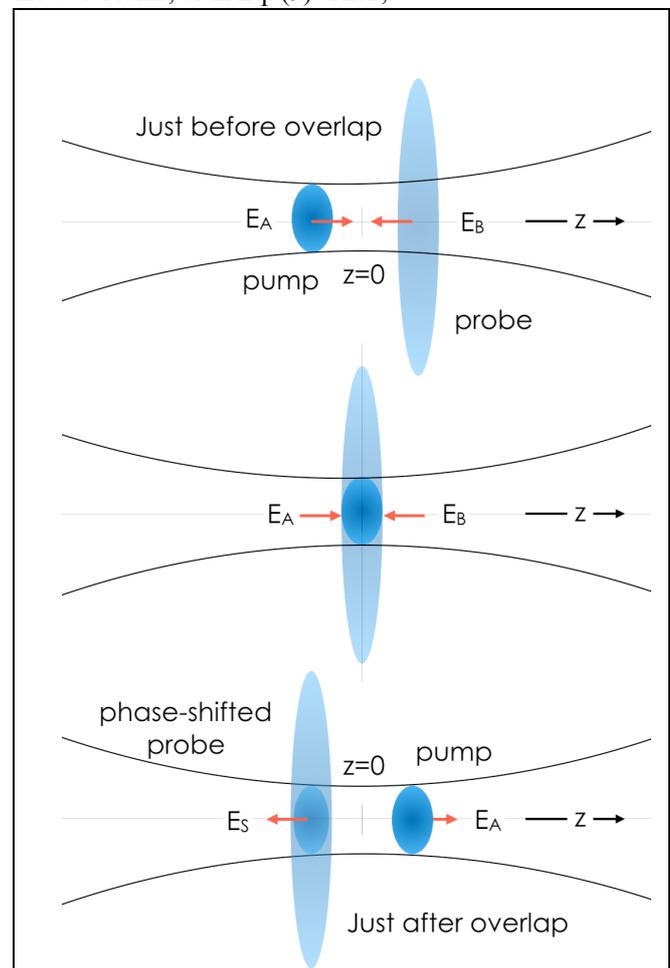

**Figure 2.-** Schematic representation of the interaction close to the focal volume. Before the interaction with the qvac perturbed by the pump beam A, the probe beam has a profile given by $E_B(x,y)$. Just after the interaction the wavefront has an added phase $E_S(x,y) = E_B(x,y)\,\phi(x,y)$. The solid curve shows the pump waist (with a short Rayleigh length). The probe has a much bigger waist and thus a much longer Rayleigh length.

$$F_S^P(\zeta) = -\frac{i}{\lambda_B} \int_0^R 2\pi\rho E_B(\rho) \exp(i\phi_P(\rho)) J_0(k_B\rho\zeta) d\rho. \quad (13)$$

We develop the parallel polarization ($E_A$ parallel to $E_B$) case, the perpendicular polarization case is similar:

$$F_S^L(\zeta) = -\frac{i}{\lambda_B} \int_0^R 2\pi\rho E_B(\rho) \exp\left(i 4 q_L \xi_L^{QED} k_B \tau_A |E_A(\rho)|^2\right) J_0(k_B\rho\zeta) d\rho \quad (14)$$

Thus, we get a function $F_S^P(\zeta)$ that is the signal as a function of the beam A's intensity and we need to compare it to the same signal without the pump beam $F_B(\zeta)$. Therefore, the qvac signal, QVS, is going to be $QVS^P(\zeta) = |F_S^P(\zeta)|^2 - |F_B(\zeta)|^2$, for the L or the T cases.

We observe that $\xi_L = q_L \xi_L^{QED} = q_L\, 6.7\, 10^{-24}\, cm^3/J$ and $\xi_T = q_T \xi_T^{QED} = q_T\, 6.7\, 10^{-24}\, cm^3/J$ are the coupling constants to compute the phase shifts at each radius of the focal spot,

$$\phi_L(\rho) = 4\xi_L I_A(\rho) k_B \tau_A = 4 q_L \xi_L^{QED} I_A(\rho) k_B \tau_A \quad (15)$$

$$\phi_T(\rho) = 7\xi_T I_A(\rho) k_B \tau_A = 7 q_T \xi_T^{QED} I_A(\rho) k_B \tau_A \quad (16)$$

with $I_A(\rho) = |E_A(\rho)|^2$ indicating the pump pulse intensity profile. The phase change depends on the pump pulse duration; therefore, we could consider that the longer the pulse the better. However, we have to observe that the experimental configuration schematically shown in Fig. 2 implies a pump that is almost spherical (its transverse width similar to its longitudinal size) or oblate (shorter longitudinally than transversally). A very long pulse tightly focused could result in a prolate pulse with a longitudinal dimension larger than the Rayleigh length, in which case not all the energy will be on the focal plane when probed by the B beam. The hypothesis to jump from Eq (9) to Eq (10) will not be valid in this long pump pulse situation, the overlapping will not occur at the waist and the qvac effect will be absolutely negligible at currently available or anticipated long-pulse facilities in the near future.

It is worth noting again that the diffraction equations considered here describe a beam with cylindrical symmetry and thus the transverse intensity profile described takes into account this cylindrical symmetry. Within this assumption, the radial profile of the Airy function at focus due to the flat top pump profile is properly accounted for. The generalization to other pulse transverse profiles (such as the rectangular shape used in some big lasers) is straightforward for the case of the pump (in the phase shift) as well as for the case of the probe.

Most of the extreme lasers are based on the Ti:Sapphire CPA technology [46, 47] that corresponds to 800 nm central wavelength and durations of the order of 30 fs or less. A reasonable choice of parameters is to consider such a pump wavelength and a probe of the same wavelength or its second harmonic. For 30 fs, 800 nm pump and probe,

$$\phi_L(\rho) = q_L\, 6.3\, 10^{-32}\, I_A(\rho)\, [W/cm^2] \quad (17)$$

$$\phi_T(\rho) = q_T\, 11.0\, 10^{-32}\, I_A(\rho)\, [W/cm^2] \quad (18)$$

where the intensity is written in W/cm². It is clear that this number is very small, and thus the effect needs a very clean and precise experiment to be seen.

*4.3 Perturbative model.*

Because the phase shift due to the qvac is going to be very small, we can determine scaling laws by considering the power expansion, $\exp(i\phi_P(\rho)) \approx 1 + i\phi_P(\rho)$, leading to this approximation,

$$F_S^L(\zeta) = -\frac{i}{\lambda_B} \int_0^R 2\pi\rho E_B(\rho) J_0(k_B\rho\zeta) d\rho + \\ + \frac{1}{\lambda_B} \int_0^R 2\pi\rho E_B(\rho) 4 q_L \xi_L^{QED} k_B \tau_A I_A(\rho) J_0(k_B\rho\zeta) d\rho \quad (19)$$

$$F_S^L(\zeta) = F_B(\zeta) + \\ + \int_0^R \rho E_B(\rho) 4 q_L \xi_L^{QED} k_B^2 \tau_A I_A(\rho) J_0(k_B\rho\zeta) d\rho \quad (20)$$

Although numerical calculations are possible with the expression given in Eq (13), the approximation in Eq. (20) is very important because it shows that in the weak limit (the only experimentally accessible limit today) the part of the scattered field $F_S$ due to the qvac scales as the intensity of the pump beam (which shows the number of scattered photons will therefore scale as the square of the pump intensity).

In the regions where $|F_B(\zeta)| \approx 0$, the signal to measure is $QVS^P(\zeta) = |F_S^P(\zeta)|^2$ and its scaling law with both lasers intensities is trivial, $QVS^L(\zeta) = Q\, q_L\, I_A^2\, I_B$, where Q is just a constant dependent on the angle. This scaling helps to design the most convenient experimental setup. Observe also that we propose to work in a region (an angle) where $|F_B(\zeta)| \approx 0$, i.e., where there are almost no photons coming from the unperturbed probe. For that reason, we propose to work with a very clean spatial Gaussian profile for the probe beam.

*4.4. Controlling the pulse delay.*

Although this is a head-on collision between two laser beams, the point where they overlap is critical. If we have precise control over the relative delay between the two beams, then we can control where they will collide. This can be used to get a characteristic signature of the qvac effect that does not appear in other sources of noise and that therefore can lead to a characteristic detection pattern. If the delay between both beams is precisely defined, $\Delta t$ then they will interact at a distance from the waist equal to $z = c\Delta t/2$ (as indicated in Fig. 3) instead of $z=0$. At this overlap point the probe beam waist will be approximately the same, but the peak intensity of the pump will be dramatically reduced, due to its tight focusing, as will the signal.

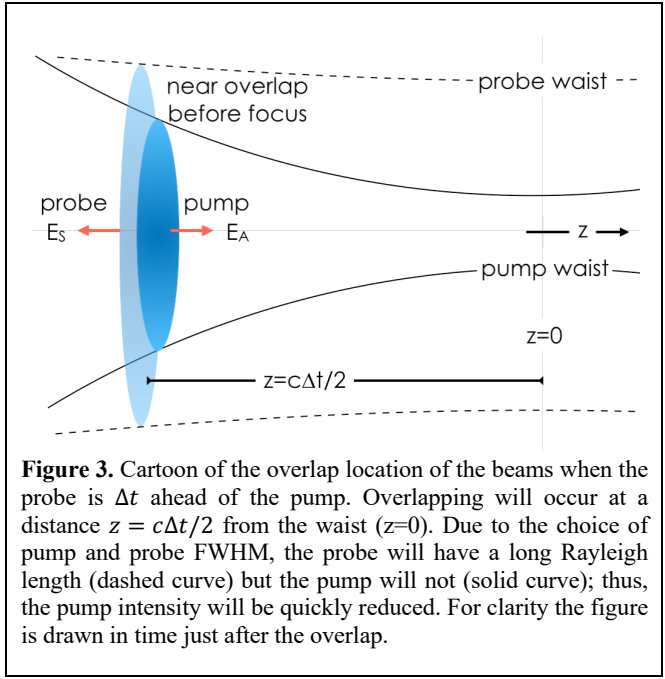

**Figure 3.** Cartoon of the overlap location of the beams when the probe is $\Delta t$ ahead of the pump. Overlapping will occur at a distance $z = c\Delta t/2$ from the waist (z=0). Due to the choice of pump and probe FWHM, the probe will have a long Rayleigh length (dashed curve) but the pump will not (solid curve); thus, the pump intensity will be quickly reduced. For clarity the figure is drawn in time just after the overlap.

The influence of this time delay can be analyzed in the perturbative model. The Airy profile, particularly its central and more intense part, can be approximated by a convenient Gaussian with the same FWHM in intensity. The waist of this Gaussian will be $w_{0A} = 0.87\,\lambda N$ (N being the f/#), having a Rayleigh length of $z_{RA} = \pi w_{0A}^2/\lambda_A$. This is only an approximation for the central part of the Airy focus, to prepare convenient scaling laws, the simulations shown in the next section are exact without this approximation. If we call $I_{0A}(z)$ the intensity of the central part (i.e., for $\rho = 0$) of the beam (axial) it is going to change with z as,

$$I_{0A}(z) = I_{0A}(0)\frac{w_{0A}^2}{w_A^2(z)} = I_{0A}(0)\frac{\pi^2 w_{0A}^4}{\pi^2 w_{0A}^4 + z^2\lambda_A^2}. \quad (21)$$

Substituting the waist by its expression in terms of the f-number, leads to:

$$I_{0A}(z) = I_{0A}(0)\frac{5.65\,\lambda_A^2 N^4}{5.65\,\lambda_A^2 N^4 + z^2} \quad (22)$$

Expressing z in terms of the jitter time, $2z = c\Delta t$ (see Fig.3), allows us to modify Eq. (21). Assuming that the relative jitter is smaller than the Rayleigh length of the probe, the qvac signal is going to be:

$$\begin{aligned}QVS^L(\zeta) &= C(\zeta)\,q_L^2\,I_{0A}^2(z)\,I_B \\ &= C(\zeta)\,q_L^2\,I_{0A}^2(0)\,I_B\frac{22.6\,\lambda_A^2 N^4}{22.6\,\lambda_A^2 N^4 + c^2(\Delta t)^2}\end{aligned} \quad (23)$$

This expression is of great experimental interest because it represents a characteristic Lorentzian shape of the dependence of the qvac signal with the pulse delay time $\Delta t$ between pump and probe where C is a constant characteristic for each polar angle $\zeta$. In an experiment where we need to get rid of an enormous noise-background, the knowledge of the signal dependence with the pulse delay is fundamental because the rest of the noise sources are not affected by a small pulse delay.

In the case when the pulse delay cannot be controlled precisely or measured for each individual shot, then the jitter pattern distribution will imply a characteristic density distribution of the qvac measurements that can also be of some use to discriminate the signal from the noise.

## 5.- Suggested experimental configuration.

There are multiple configurations of pump and probe beams. Among them we are going to propose one that has a great potential for a future experiment.

### 5.1 Pump beam (beam A).

The pump intensity needs to be as high as possible, from a multipetawatt laser. In most CPA lasers an efficient extraction of the pump energy stored in the last amplifiers causes a saturation and a profile close to a flat top. Thus, we consider an Airy profile (Fourier transform of a flat-top with the aperture of the focusing OAP acting approximately as a circular aperture). This beam is going to have a wide diffraction pattern, but is propagating away from the detector, so it is not important. The key feature is just the intensity in this case. At the focal plane the pumping field intensity will take the form of the well know Airy focal spot [45],

$$I_A(\rho) = I_{0A}\,2\frac{\lambda_A N}{\pi\rho}J_1\!\left(\frac{\pi\rho}{\lambda_A N}\right) \quad (24)$$

with N=f/D, f the effective focal length of the parabolic mirror and D the mirror diameter. The f-number of the OAP is very important because it determines the angular region of space where the pump laser focusing system prevents the convenient positioning of a detector. Therefore, the detection angle has to be slightly bigger than $(D/2)/f$. The full width at half maximum, FWHM, of the pump beam is given by $\text{FWHM} = 3.24\,\lambda_A\,N/\pi \approx \lambda_A\,N$.

*5.2 Probe beam (beam B).*

We consider a high-quality probe beam with minimum diffraction, i.e., a Gaussian beam. We want to detect the few photons scattered by the qvac mechanism and thus to maximize signal to noise, the cleaner the B pulse profile the lower the noise level generated by this beam. However, there are many other sources of noise that will be discussed later in this manuscript.

The probe beam B will be,

$$E_B(\rho) = E_{0B}\exp(-\rho^2/w_B^2) \quad (25)$$

which helps to reduce the scattered light from the probe. However, the scattered light from the pump is going to be difficult to avoid. As mentioned earlier, we propose instead of comparing the QVS with and without the pump to compare the QVS with and without pump-probe synchronization. In the first case we get the true signal and in the second case only the noise from the pump, provided that the two lasers have a time offset of a few ps that prevents them to overlap at the focus (if the overlapping is beyond the Rayleigh zone the qvac effect will not play any role. So experimentally it is important to have a good time synchronization system.

**6.- Results.**

For this effect to be measurable, the intensity of the pump beam must be sufficient to overcome background noise. To that end, one wants to use and OAP with the shortest practical f-number. In principle very short f-numbers are possible. Today's intensity record of $1.1\ 10^{23}$ W/cm$^2$ [48] was achieved with f/1.1. A focus that tight, however, requires the vector character of the field to be taken into account, which will mix parallel and perpendicular components and thus blurring the result we are looking for. A safe compromise would be f/3 that we estimate will generate an intensity of $10^{24}$ W/cm$^2$ with a 10 PW laser.

*6.1 Number of photons of the probe beam on axis.*

Just to get a sense of how many photons per solid angle we have in the probe beam we can analytically calculate the propagation (nonperturbed propagation) of the probe beam close to the z axis.

We calculate the intensity on axis for a small solid angle, 1 deg$^2$, centered at the z axis, we call this axial intensity along z, $I_{axial}(z)$ and the solid angle $d\Omega$. With $w_B$ as the waist of beam B and $I_B$ its the intensity at the waist, the probe intensity along the axis will be expressed as,

$$I_{axial}(z) = I_B\left(\frac{\pi w_B^2}{z\lambda_B}\right)^2 \quad (26)$$

and the surface corresponding to this solid angle is $z^2 d\Omega$. Therefore,

$$I_{axial}(z)z^2\,d\Omega = I_B\left(\frac{\pi w_B^2}{\lambda_B}\right)^2 d\Omega \quad (27)$$

The fluence will be this expression multiplied by $\tau_B$. Each Joule corresponds to $4\ 10^{18}$ photons, at 800 nm. Thus, the number of photons within 1 deg$^2$ solid angle will be

$$n_{photons} = 4\ 10^{18}\frac{\text{photons}}{\text{Joule}}I_B\,\tau_B\frac{1}{4}k_B^2\,w_B^4\,d\Omega \quad (28)$$

In the case of $I_B=10^{20}$ W/cm$^2$, that is a very demanding beam for a good Gaussian shape and that requires a lot of filtering, (and for 30 fs pulse duration). Observe that 0.000305 sr correspond to 1 deg$^2$. Therefore, for $w_B$=20 wavelengths=16 μm this corresponds to $n_{ph} = 3.7\ 10^{19}$ photons.

The number $3.7\ 10^{19}$ photons per shot, for $I_B=10^{20}$ W/cm$^2$, arriving to the detector (on axis) for $w_B$=16 μm (20 wavelengths), for 30 fs pulse duration and a detector on axis covering 1 deg$^2$ solid angle, is going to be considered as a realistic reference for the comparisons presented in this paper. The scattering for other intensities of the probe beam can easily be calculate because of the linear scaling of the number of photons with $I_B$.

*6.2 Numerical results for f/3.*

As mentioned above, we consider that a reasonable f-number to keep the scalar description and thus, a clear differentiation between parallel and perpendicular is N=3. Results for higher values of the f-number are interesting but they imply an even more powerful laser pulse. While smaller by a factor of $(7/4)^2$, in this paper we present results for the parallel case. The perpendicular case would lead to a signal three times stronger.

The scattered field Fraunhofer diffraction profile for several values of the pump intensity has been computed based on Eq. (12). In particular we considered three values of the intensity, $I_A=10^{23}$ W/cm$^2$, in agreement with today's experimental achievement [48]. We also considered $I_A=10^{24}$ W/cm$^2$, as a value reachable in the present decade with foreseeable multipetawatt lasers. Also, an extreme case of $I_A=10^{26}$ W/cm$^2$ is depicted. This huge field does not seem possible with the technology currently under consideration, but we include it because it shows the validity of the perturbative model even at this extreme value. We must point out that at $10^{26}$ W/cm$^2$, radiation reaction would typically be of concern. However, we do not see this as a limitation here because we will need to work at record low vacuums and nearly all free electrons will be expelled from the focal volume by the beams (acceleration and ponderomotive repulsion) long before the peak intensity is reached.

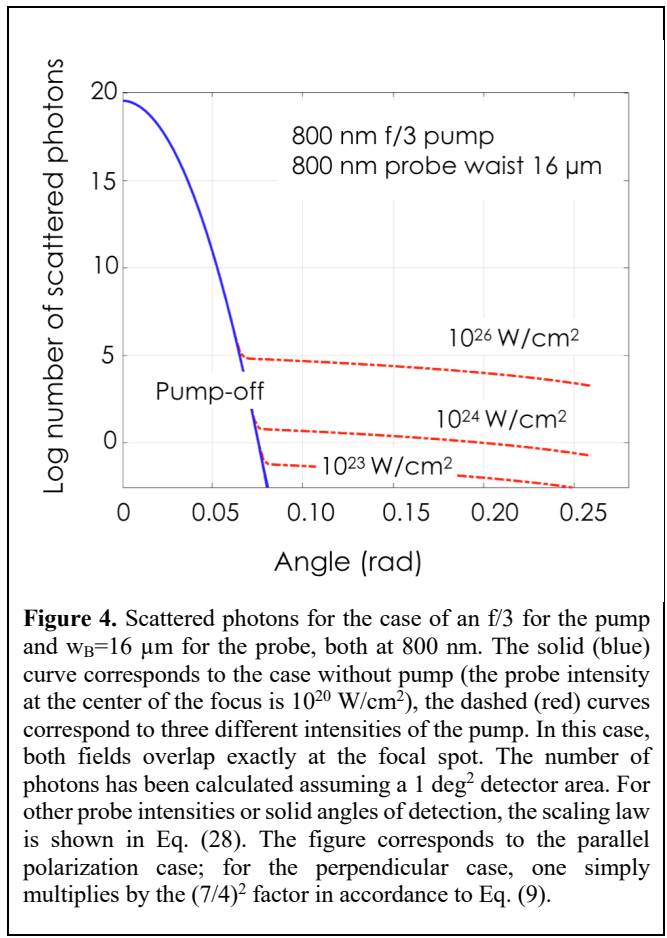

**Figure 4.** Scattered photons for the case of an f/3 for the pump and $w_B$=16 μm for the probe, both at 800 nm. The solid (blue) curve corresponds to the case without pump (the probe intensity at the center of the focus is $10^{20}$ W/cm$^2$), the dashed (red) curves correspond to three different intensities of the pump. In this case, both fields overlap exactly at the focal spot. The number of photons has been calculated assuming a 1 deg$^2$ detector area. For other probe intensities or solid angles of detection, the scaling law is shown in Eq. (28). The figure corresponds to the parallel polarization case; for the perpendicular case, one simply multiplies by the (7/4)$^2$ factor in accordance to Eq. (9).

The numerical example shown in Fig. 4 indicates the number of scattered photons of the pump beam as a function of the diffraction angle. We note that this, and other numerical plots in this paper, assume (1) an initial number of photons corresponding to a probe intensity of $10^{20}$ W/cm$^2$ and (2) a detector with unit quantum efficiency and a collection solid angle of 1 deg$^2$. Larger solid angles or lower quantum efficiencies can be determined with a multiplicative factor. The value of 1 deg$^2$ solid angle is just a differential solid angle representative of a small detector element which could be used in a real experiment.

The solid curve (blue) indicates the diffraction pattern of the unperturbed probe (i.e., with the pump pulse off). Because the probe has a wide Gaussian waist, the number of scattered photons (due to diffraction) falls very quickly with the angle. Observe that the plots vertical scale is logarithmic (log$_{10}$) so the Gaussian shape maps into a parabola. The dashed curves (red) indicate the qvac scattering (in the region where this scattering is not hidden by the unperturbed pattern) for the three characteristic values of the pump intensity. Dashed lines in this and the rest of figures are hidden below the solid one when the nonlinear scattering is irrelevant. Figure 5 is a zoom of Fig. 4 evidencing (dotted line) the one photon per shot line (for $10^{20}$ W/cm$^2$ probe intensity and 1 deg$^2$ detector area) due to qvac scattering.

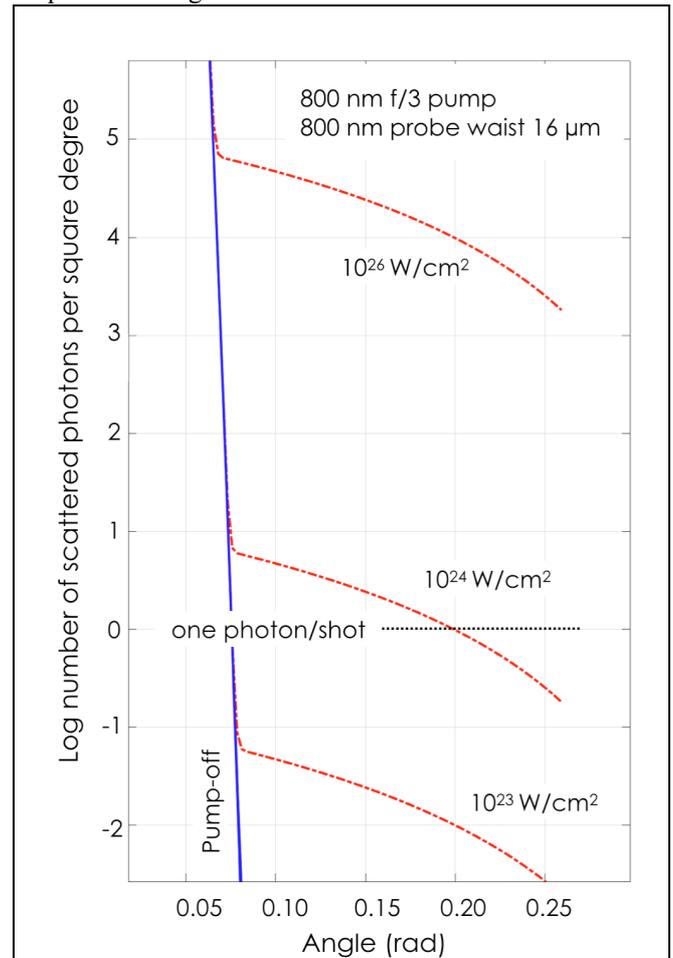

**Figure 5.** Zoom of Fig. 4 to show more clearly the one-photon per shot line. For $I_{AO}$= $10^{24}$ W/cm$^2$ there is a range of scattering angles where the number of scattered photons due to the qvac lies above the one-photon per shot line. The figure corresponds to the parallel polarization case.

This family of photon-photon scattering experiments performed well before the onset of pair creation represents a clean window to study the essence of the qvac. When pairs are possible many new effects complicate the experimental scenario (electron radiation, electron positron showers, … and other things some of which are likely unexpected). Our working region, at the onset of the qvac refractive index modification without real particle creation is a very delicate experiment but provides novel measurements of QED – a direct measure of the Lagrangian. There are a number of sources of noise and potential systematic errors that need attention, some of which we discuss below.

*6.3 Where to place the detector?*

The previous plots show that it is best to detect the qvac scattered photons as close as possible to the axis. Using $w_B$=16 $\mu$m (for 800 nm probe), corresponds to a half angle of the Gaussian equal to $1/20\pi$ rad (0.016 rad). This is for the $1/e$ decay. However, in order for the unperturbed contribution due to diffraction (solid blue curve in the figures of Section 6) to be negligible, it is better to observe at angles several times larger than the $1/e$ decay. A reasonable lower limit is the typical three times this value, $1/e^3$, corresponding to 0.048 rad. Unfortunately, this is not the only limitation. The pump beam with an f-number N=3 implies an angle from the axis of $1/6$ rad = 0.17 rad.

The simulations would suggest placing the detector between 0.08 and 0.10 rad. However, this would mean that the pump OAP would interfere with the detection (see Fig. 6). The minimum angle is 0.17 rad. Another choice could be to drill a small hole at the off-axis-parabola mirror used to focus the pump beam and work at angles smaller that 0.10 rad, as shown in Fig. 6. However, we do not consider this a realistic option due to scattering induced by the hole and the difficulty in filtering the pump radiation at the same wavelength.

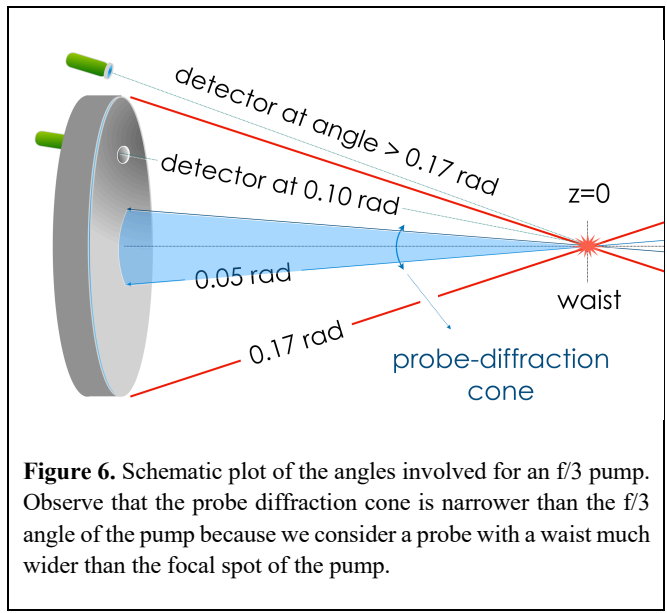

**Figure 6.** Schematic plot of the angles involved for an f/3 pump. Observe that the probe diffraction cone is narrower than the f/3 angle of the pump because we consider a probe with a waist much wider than the focal spot of the pump.

*6.4 Results for second harmonic probe.*

A third possibility, often considered in the literature, is a second harmonic probe. Our purpose is just analyze the effect of a second harmonic probe. The losses to generate the second harmonic have not been considered here. Results for that case, with f/3, are shown in Figs. 7 and 8. They show that the second harmonic probe reduces the angle and thus increases the signal but at the cost of fewer photons (each with twice the energy). However, the one photon per shot line, depicted in Fig 7, now cuts the $10^{24}$ W/cm$^2$ pump red line at an angle smaller than the pump aperture cone. Thus, the observation line intercepts the pump focusing mirror. In this case, the hole to be drilled in the pump OAP (see Fig. 6) could be an option, with the use of dichroic and color-glass filters to block the 800 nm scattered light. A possible technical advantage of this fundamental + second harmonic configuration is that the dichroic mirrors might enable a perfect counterpropagating beams arrangement.

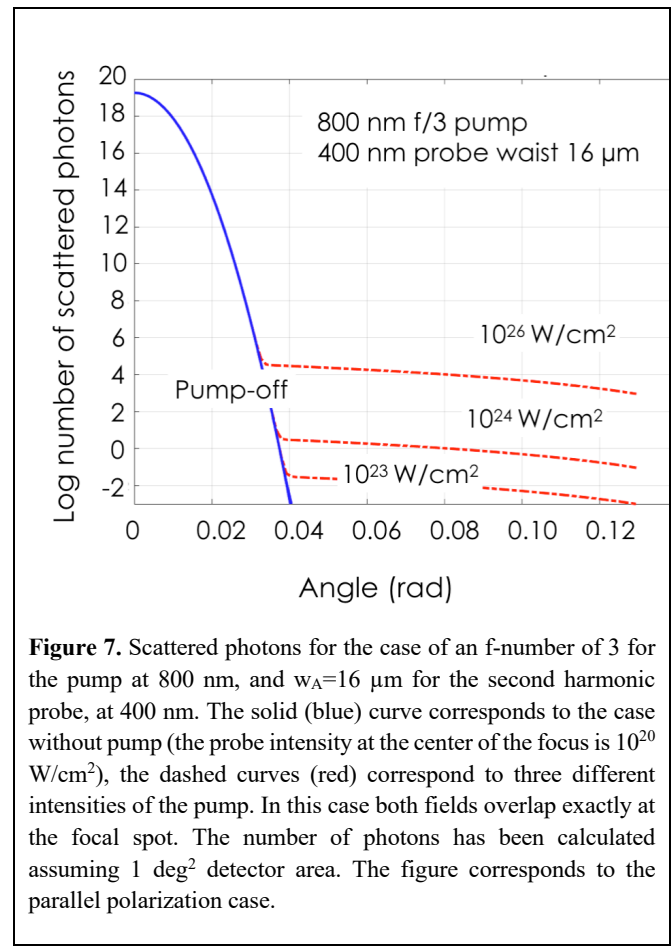

**Figure 7.** Scattered photons for the case of an f-number of 3 for the pump at 800 nm, and $w_A$=16 $\mu$m for the second harmonic probe, at 400 nm. The solid (blue) curve corresponds to the case without pump (the probe intensity at the center of the focus is $10^{20}$ W/cm$^2$), the dashed curves (red) correspond to three different intensities of the pump. In this case both fields overlap exactly at the focal spot. The number of photons has been calculated assuming 1 deg$^2$ detector area. The figure corresponds to the parallel polarization case.

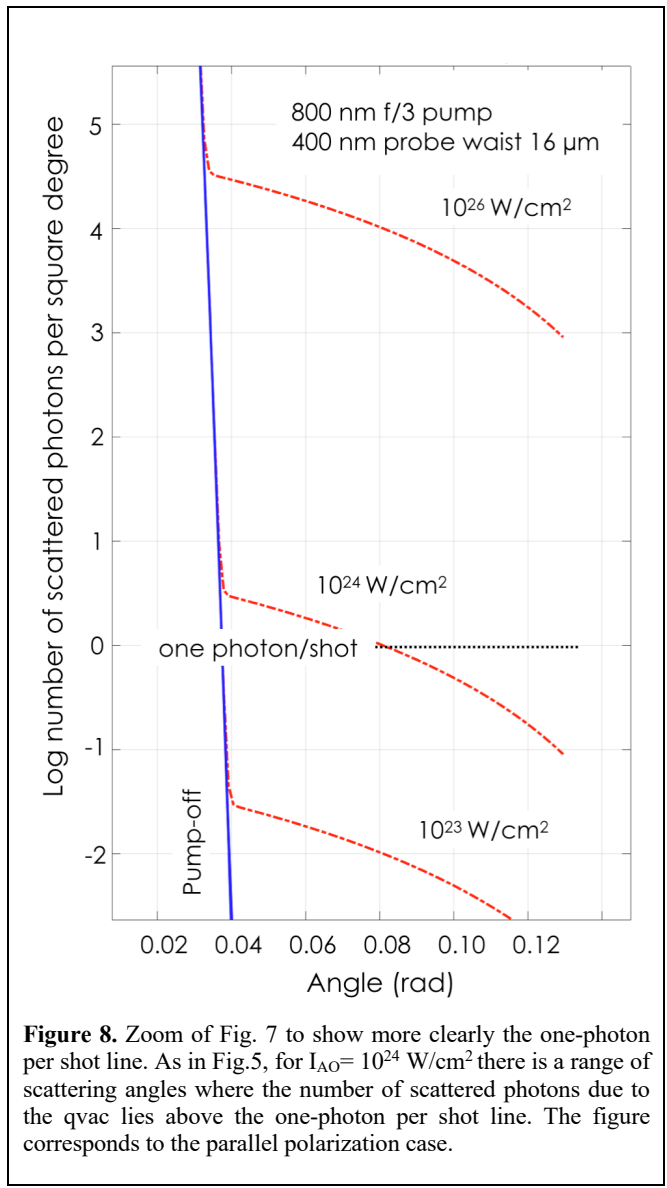

**Figure 8.** Zoom of Fig. 7 to show more clearly the one-photon per shot line. As in Fig.5, for $I_{AO}=10^{24}$ W/cm² there is a range of scattering angles where the number of scattered photons due to the qvac lies above the one-photon per shot line. The figure corresponds to the parallel polarization case.

### 6.5 Effect of the probe beam waist.

Up to now we have stated that it is best to have a probe with a large waist to prevent ordinary diffraction into the detection region of interest for measurement. We will now consider this in more detail. We also will show there are practical limits. We can show this by comparing our result for $w_B=16$ μm (Fig. 4) with a probe with larger and smaller waists, but at the same probe intensity.

Figure 9 is equivalent to Fig. 4 but for $w_B=24$ μm. The solid blue curve in Fig. 9 shows the ordinary diffraction of the probe that is much narrower due to the larger waist. The dashed red curve shows the qvac effect for the chosen three pump intensities. The values in the region not hidden under the linear diffraction component are the same for Figs. 4 and 9, showing that the effect of the waist is simply to unveil a region of ζ angles, however this region may enter into conflict with the pump beam size considerations discussed above due to the f/3 OAP. At the same time, even though a waist of 24 μm has a diffraction smaller than the original case, to achieve the same intensity the power has to be 9/4th larger for this larger waist (the required power scales as the waist squared).

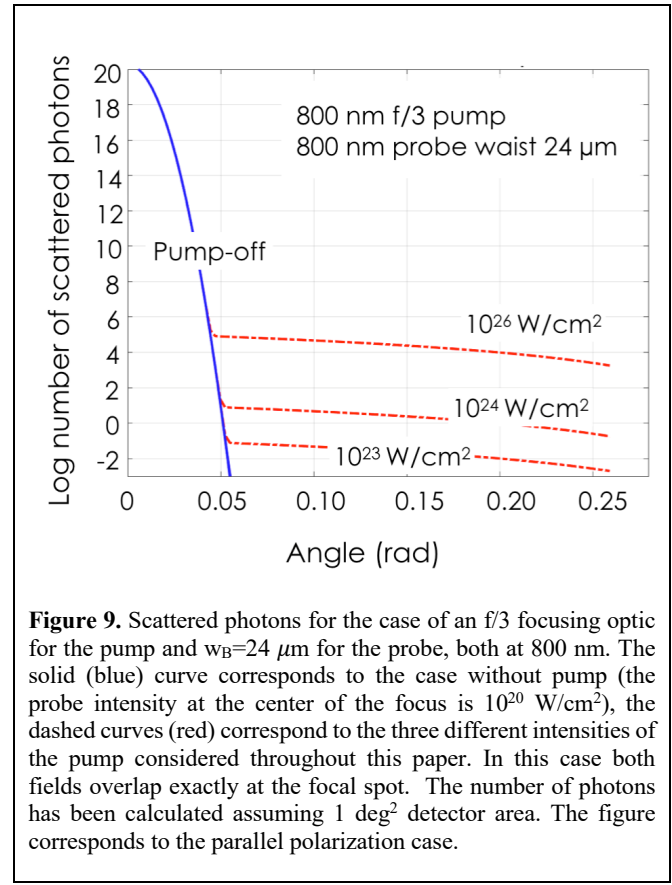

**Figure 9.** Scattered photons for the case of an f/3 focusing optic for the pump and $w_B=24$ μm for the probe, both at 800 nm. The solid (blue) curve corresponds to the case without pump (the probe intensity at the center of the focus is $10^{20}$ W/cm²), the dashed curves (red) correspond to the three different intensities of the pump considered throughout this paper. In this case both fields overlap exactly at the focal spot. The number of photons has been calculated assuming 1 deg² detector area. The figure corresponds to the parallel polarization case.

While a large $w_B$ waist implies more energy per pulse for the probe beam (for the same probe intensity at the center of the focus, $10^{20}$ W/cm²), it would be less demanding on probe pulse energy to consider a smaller waist. Figure 10 shows the result for the same conditions as Fig. 4 but for a probe waist of only 8 μm, which will require one fourth the energy per shot compared to Fig. 4. In spite of the fact that the FWHM of the probe is larger than that of the pump (so the pulses overlap sufficiently for the qvac effect to occur), the probe linear diffraction pattern is so broad that it hides the interesting region of nonlinear scattering. Consequently, a probe beam of ~ 16 μm is near the sweet spot for these experiments.

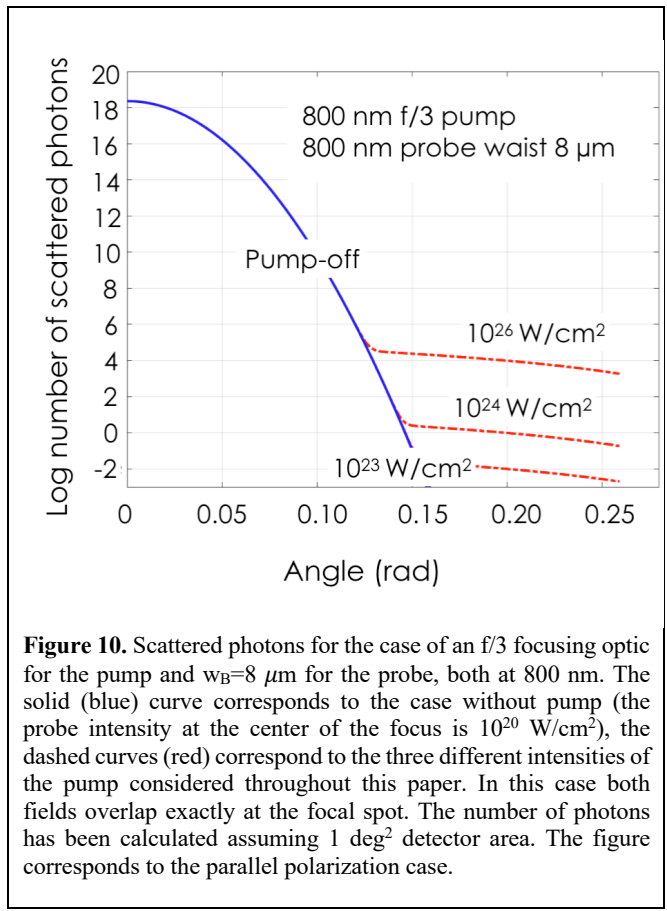

**Figure 10.** Scattered photons for the case of an f/3 focusing optic for the pump and $w_B=8$ μm for the probe, both at 800 nm. The solid (blue) curve corresponds to the case without pump (the probe intensity at the center of the focus is $10^{20}$ W/cm$^2$), the dashed curves (red) correspond to the three different intensities of the pump considered throughout this paper. In this case both fields overlap exactly at the focal spot. The number of photons has been calculated assuming 1 deg$^2$ detector area. The figure corresponds to the parallel polarization case.

*6.6 Results for f/10.*

We also present results for a larger f-number for the pump beam. Observe, however, that the increase of the f-number represents an increase of the FWHM for the pump. The results are shown in Fig. 11 and its zoom for the interesting region in Fig. 12. Again, a higher power than the case in Fig. 4 is needed to keep the same peak intensity. Figures 11 and 12 show, as expected, that as the pump waist increases the qvac diffraction it induces is smaller.

From that result we can conclude that the smaller the pump waist the better. The qvac diffraction angle is inversely proportional to the pump waist. However, this has a limit. As stated earlier, for extremely short f numbers (e.g., f/1.1 [48]), the longitudinal field at the waist is too strong to consider a pure parallel or perpendicular case and thus the $q_L$ and $q_T$ parameters cannot be independently measured. We conclude that an f/3 focusing optic for the pump appears to be a promising region to identify the $q_L$ and $q_T$ parameters separately. Such f/3 focusing for the pump allows relatively high intensities and small diffraction angles with a negligible influence of the longitudinal component, and thus with a relatively clean separation between parallel and perpendicular polarizations.

Figures 11 and 12 show the result for a f/10 pump and the 16 μm probe waist. Increasing both the pump f number (i.e. its waist) and the probe waist (keeping its TEM$_{00}$ shape) and keeping at the same time the pump and the probe intensities would lead to a very convenient scenario for improved measurements of the Lagrangian coefficients $q_L$ and $q_T$. However, this would need much more energy per shot. For that reason, we suggest that f/3 is a convenient starting point.

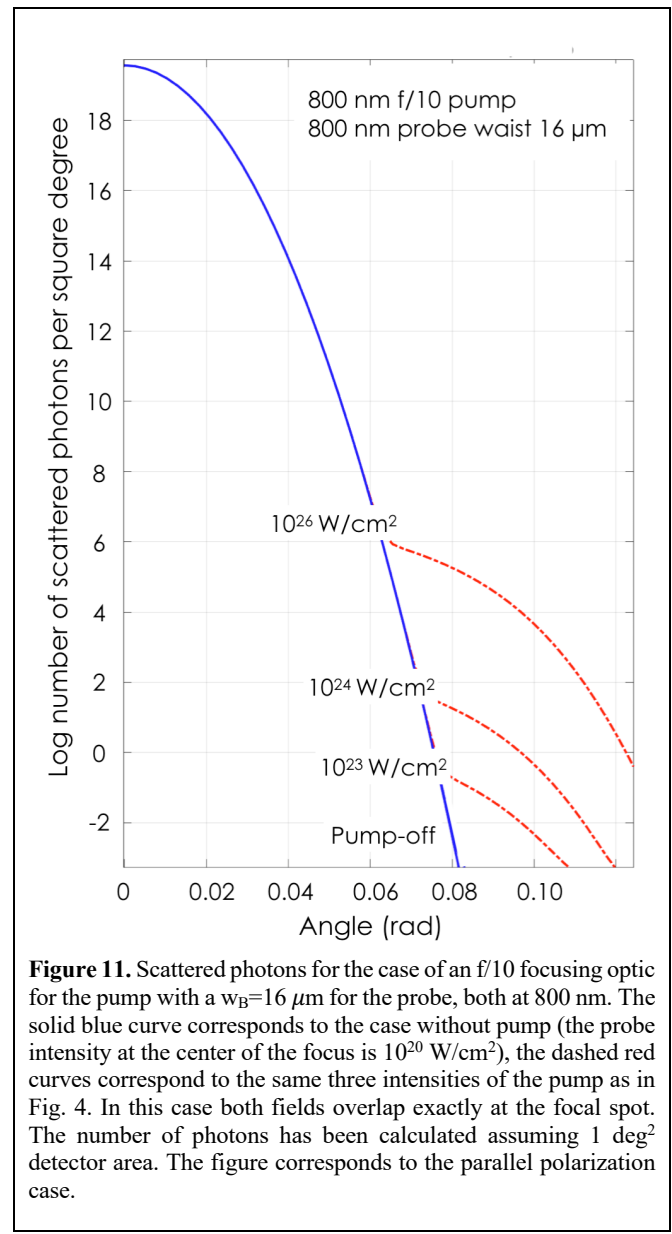

**Figure 11.** Scattered photons for the case of an f/10 focusing optic for the pump with a $w_B=16$ μm for the probe, both at 800 nm. The solid blue curve corresponds to the case without pump (the probe intensity at the center of the focus is $10^{20}$ W/cm$^2$), the dashed red curves correspond to the same three intensities of the pump as in Fig. 4. In this case both fields overlap exactly at the focal spot. The number of photons has been calculated assuming 1 deg$^2$ detector area. The figure corresponds to the parallel polarization case.

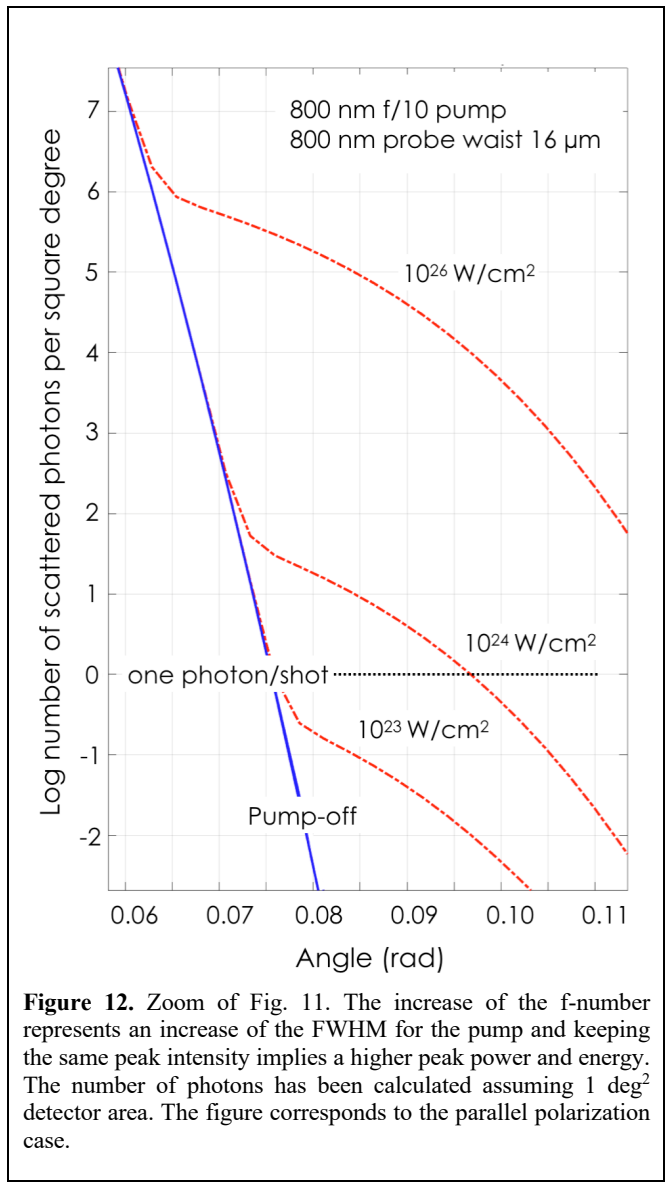

**Figure 12.** Zoom of Fig. 11. The increase of the f-number represents an increase of the FWHM for the pump and keeping the same peak intensity implies a higher peak power and energy. The number of photons has been calculated assuming 1 deg$^2$ detector area. The figure corresponds to the parallel polarization case.

## 7.- Analysis of a future experimental scenario.

The advantage of the wide-waist Gaussian probe is that the number of scattered photons will be extremely small, and this is key for the qvac signal $\text{QVS}^P(\zeta) = |F_S^P(\zeta)|^2 - |F_B(\zeta)|^2$, which in the region of observation is $\text{QVS}^P(\zeta) = |F_S^P(\zeta)|^2$. However, this signal is going to be very tiny and any source of noise not accounted for can impair observation and the desired precision measurement of the Lagrangian.

### 7.1 Sources of noise.

This experiment has an intrinsic contradiction, on the one hand we need to measure a very weak signal, and on the other we have to work with the world's most extreme lasers with huge numbers of photons. The main noise sources for these experiments come from the following phenomena:

1) Scattered light from the walls, apertures and parabolas, both from the pump and from the probe pulses (particularly if they are of the same frequency).

2) Thomson scattering from electrons coming from residual gas. This includes regular Thomson scattering over the full volume where gases are ionized (I > $10^{14}$ W/cm$^2$) and relativistic Thomson scattering (RTS) from electrons within several Rayleigh ranges of the focus.

3) Thomson scattering from the residual ions in the focal volume.

4) Rayleigh scattering from residual gas along the beam paths in the vacuum chamber.

5) Recombination radiation from the ionized residual gas or fluorescence radiation from optical components.

### 7.2 Strategies for a successful experiment.

There are a number of strategies that will be required to enhance light detection and mitigate the effects of noise sources. One benchmark that will be used for some of the best conditions achievable is the state-of-the-art techniques used in the current LIGO facilities [49].

**Light Detection.** There are two choices of photodetector elements which can be employed, either a photocathode surface (Streak camera, photomultiplier or gated intensified imager) or a semiconductor detector (silicon or other). For Photocathode detectors quantum efficiencies of ~15% (40%) can be obtained at 800 nm (400 nm) with GaAs(Cs) (S20) photocathodes [50]. For silicon semiconductor detectors quantum efficiencies of around 80% (40%) are achievable at 800 nm (400 nm) [51]. Also, one can take advantage of avalanche enhancement of signals in semiconductor detectors. We assume that the collected light can be reimaged to a small spot, to the order of 25 μm, onto a small low-noise detector element if a small detector is used or a single pixel if an imaging detector is employed. Also, it is assumed that using a fast time response or short time gated detector (< 1 ns), coincident with the expected pulse that the average photon shot noise or dark current noise in the detector would be less than 0.01 photons equivalent noise per pulse (<10$^7$ photons per second shot noise for a 1 Hz laser system). At present, using a small high speed Si photodiode detector at 80% quantum efficiency appears most promising. It is assumed that any stray background x-ray radiation from the few electrons accelerated in the extended interaction region will be sparse (estimated number of source electrons will be less than $10^6$, see discussion below) and the probability of an x-ray hitting the very small detector volume will be negligible. Thus, we will assume the detector shot noise is less than 0.01 photons per shot. In this case the major source of noise will be due to scattered light collected by the detector.

**Scattered Light**. A number of approaches can be used to minimize scattered light to state-of-the-art levels, which could include the following:

1. Superpolished optical surfaces should be used, with RMS roughness of less than ~ 0.5 nm, for the final turning mirrors and last few beamline mirrors. Less than ~$10^{-6}$ energy scattering per surface would be expected into a $\cos(\theta)$ type distribution from such surfaces.

2. A physical aperture of approximately 4 mm diameter at the focal plane would be required to block direct line of sight scattered light from the final focusing mirror to the detector, thus requiring multi surface scattering to arrive at the detector. For this reason, simulations presented above consider such an aperture in the system.

3. A confocal imaging system for the detector system restricted to approximately a 3000 μm long by 50 μm diameter region of the focal interaction region.

4. Narrow bandpass filters on the detector to block out-of-band recombination radiation, fluorescence and Doppler shifted RTS. A single bandpass filter would achieve a rejection ratio of better than $10^{-3}$ of these background noise sources and, if necessary, multiple bandpass filters could be employed in series to increase the rejection of out of band noise, with a modest reduction of detected signal levels.

5. An ultrashort optical shutter of the order of 10 ps to discriminate against scattered light from stray optical surfaces arriving at different times than the real signal. Such a shutter could be in the form of a Streak camera, gated intensified imager or an ultrafast Kerr shutter [52]. It is expected that state-of-the-art systems can achieve $10^{-4}$ rejection of out-of-gate signals.

6. Ultrahigh vacuum in the interaction region of $10^{-9}$ mbar to reduce the background particle density as much as possible. Such a vacuum in a large-scale system has been achieved in the current generation of LIGO detectors [49].

7. Cleaning of the majority of the electrons from the interaction volume by the ponderomotive force of the high intensity laser pulses themselves or by use of a lower intensity, $10^{20}$ W/cm$^2$, cleaning pulse several picoseconds prior to the main pulse. See, for example, Figs. 9 and 10 and associated text in paper by Quesnel and Mora [53].

8. Characterizing the scattered light by taking reference shots with the probe delayed by up to 1 ps. All scattered radiation sources would remain identical in these reference shots while the signal would go to zero with picosecond delays between the pump and probe pulses. A full analysis of stray scattered light from all sources would require a detailed assessment of each source, along with transport and coupling of this scattered light to the final detector element. Here, we only give a brief, order of magnitude assessment, which reveals the feasibility of a successful measurement even at the $10^{23}$ W/cm$^2$ pump level.

One critical feature would be the ability to time-gate the detection on a picosecond time scale, either with a picosecond streak camera or with a short pulse driven optical Kerr shutter [52]. The Kerr shutter can be driven at the second or third harmonic of the laser pulse so that any scattered light that it generates can be filtered using bandpass filters before the detector. A gate width of 10 ps will be assumed for the shutter. Using high-quality polarizers, we expect the shutter to have a leakage level of below $10^{-4}$.

A second critical feature would be the implementation of confocal imaging that will limit the field of view of the detection system to an effective aperture of 50 μm diameter about the interaction focal volume, significantly limiting all radiation generated along the beam paths away from the focal volume from reaching the detector. Combined with the 10 ps gating, this would give an observed volume of 50 x 50 x 3000 μm or 7.5 $10^{-12}$ m$^3$, reducing scattered light from the chamber walls and other surfaces in the vacuum chamber by orders of magnitude.

A third critical feature is achieving ultrahigh vacuum conditions (of the order of $10^{-9}$ mbar), which corresponds to a density of 2.5 $10^{13}$ molecules/m$^3$. At these pressures, the residual gas is primarily hydrogen molecules. This leads to approximately 187.5 molecules or 375 liberated electrons and protons in this viewed volume. We anticipate that almost all of the electrons will be expelled from the volume by the extreme ponderomotive force established by the pump and probe beams.

To get a sense of the radiation noise resulting from these 375 electrons, we will first consider the RTS signal from the electrons and nonrelativistic Thomson scattering from the residual 375 protons within the viewed volume. While the electrons will experience enhanced RTS [54] the enhanced scattering goes into harmonics and the scattering per harmonic is of the order of the linear Thomson scattering coefficient of 6.65 $10^{-29}$ m$^2$ or 7.9 $10^{-30}$ m$^2$ sr$^{-1}$. In addition, the scattering is Doppler shifted. We will use the linear scattering coefficient as an estimate of the expected scattering within the spectral band detected. The total cross section of 375 electrons is thus of the order of 3.0 $10^{-27}$ m$^2$ sr$^{-1}$. Assuming a total of 11 PW beam power in 30 fs pulse, or 330 J through the 50 μm diameter interaction volume, the number of scattered 800 nm photons would be 2015 ph/sr or 0.6 ph/deg$^2$. However, we expect the ponderomotive cleaning to expel close to 100% of the electrons from the viewed volume at around 1% of the peak intensity (starting at the relativistic intensity of $10^{18}$ W cm$^2$). Assuming that electrons observe an average intensity of around 1% of the peak intensity as they are being swept out, we would expect around 1% of the peak evaluated above, giving an estimated RTS signal from electrons to be less than 0.01 ph/deg$^2$. It is expected that by use of a ponderomotive cleaning prepulse, if necessary, at least 90% and up to 99% of the electrons can be expelled from the interaction region prior to the main pulse. Assuming expulsion of electrons with energies of the order of 100 keV, they are expected to leave the interaction region laterally at angles of the order of 70° [55] and in 1 picosecond will travel 170 microns radially outwards clearing the interaction region. The Thomson scattering cross-section for protons is $10^6$ times less than that for electrons and thus the residual protons will lead to a scattered signal of the order of 8 $10^{-7}$ ph/deg$^2$ that can be ignored. While there will be Thomson scattering from electrons along the path of the incoming probe and outgoing pump beams outside the interaction volume, the confocal viewing system only images a column in space at an angle of approximately 0.1 rad relative to the laser axis. Thus, radiation originating outside 3000 $\mu$m viewing region will not be seen by the viewing system. Thus, we can ignore any additional Thomson scattering in the incoming and exiting beams.

Rayleigh scattering (scattering cross section ~2 $10^{-32}$ m$^2$ per hydrogen molecule [56]) will occur over all the laser beam paths in the background gas regions outside of the ionization and breakdown region leading to a background scattered-light signal in the chamber. Assuming 2.5 $10^{13}$ molecules/m$^3$ chamber pressure and 10 m path lengths for 330 J of laser light would lead to scattered light fraction of 2 $10^{-18}$ of the incident light energy or ~ 6600 scattered photons, which could easily be blocked from detection by the confocal imaging and time gating.

An important scattering contribution would come from the scattering from the final parabola surface themselves. We assume, by using super polished mirror surfaces [49] we will be able to limit the scattering to ~$10^{-6}$ of the incident light. This will produce a huge number of scattered photons, ~1.3 $10^{15}$. While the scattering would not enter the viewing system directly it would enter by diffraction or scattering off the edge of the 4 mm aperture that would be synchronous with the time gate window for both the pump beam and probe beam photons. Assuming the primary scattering from the mirror surfaces occurs into a solid angle of approximately 1 sr we estimate the photon flux due to the pump mirror at 3 m and probe laser mirror at 8 m to be ~1.33 $10^{14}$ ph/m$^2$. Assuming diffraction of the probe or scattering of the pump at the edge of the aperture has an effective cross-section of approximately one wavelength effective width around the full circumference into a solid angle of 1 sr leads to diffraction scattering of the initial flux into the entrance aperture of the confocal imaging system with a flux of approximately 1.7 $10^6$ ph/sr or 510 ph/deg$^2$. Direct viewing of these edge scattered photons would be blocked by the confocal imaging system but would scatter off the optical surfaces of the confocal imaging system. Assuming secondary scattering of these photons into the detector element can be limited to the order of $10^{-4}$ this would contribute a noise signal of the order of 0.05 photons per shot.

Finally, very efficient beam dumps will be required for both laser beams to reduce ambient light scattering background. While optical time gating will reduce this light, it will not eliminate it due to the ~$10^{-4}$ optical leakage through a typical time gate. It is expected that backscatter from the exiting optical beams would be limited of the order of $10^{-6}$. This would either be off of the superpolished opposite mirrors before reflecting far down the input optical beam paths or by using superpolished Brewster angle absorber plates as beam dumps. This will lead to an additional flux of 1.3 $10^{15}$ photons scattering in the chamber. Together with the equal amount of light scattered from the primary mirror surfaces a total of 2.6 $10^{15}$ photons will initially be scattered inside the chamber. Assuming a spherical 10 m diameter scale size chamber and optical absorbing surfaces with 1% scattering per bounce (high quality optical black surface) would lead to an average ambient re-emitted background fluence of 8.2 $10^{10}$ ph/m$^2$ after one randomizing bounce. A 1 deg$^2$ viewing cone from the center of the chamber would observe an area of 7.6 $10^{-3}$ m$^2$ on the opposite wall at 5 m, which emits 6.2 $10^8$ photons into a Lambertian cone angle of $\pi$ sr. The spatial filter in the detector system will only collect the light that starts from the wall and passes through the imaged 50-$\mu$m diameter observation disk at focus. The solid angle of this 50-$\mu$m diameter aperture region from the wall is 7.8 $10^{-11}$ sr, leading to a background signal of 0.02 ph/shot passing through the confocal filter. Additional baffling, such as a beam dump could be arranged in the direct line of sight of the detector viewing system to reduce this line of sight scattered light signal further to less than 0.01 ph/shot. With optical gating this is further reduced by approximately 4 orders of magnitude allowing the accommodation of higher scattering levels from the beam dump and optical components.

In summary, for the given conditions and a 1 deg$^2$ detector collection aperture the major contributions to photon noise per shot are detector shot noise ~0.01 ph/shot, residual Thomson scattering at ~0.01 ph/shot, re-scattering from the 4 mm aperture edge of ~0.05 ph/shot, leakage of background scattered light scattered into the detector system and leaking through the time gating of < 0.01ph/shot. Thus, overall, it is expected that it is possible to reach a background noise rate of ~0.1 ph/shot with all the above techniques. It is possible that the effect of the edge scattering from the 4-mm aperture can be reduced further by using superpolished optics in the detector collection system to minimize re-scattering of off-axis light entering the system. Thus, an ultimate background noise level of the order of 0.01 ph/shot may be achievable. Approximately a factor of 10 reduction in scattered light signal could be achieved if the probe pulse were second harmonic since 89% of the scattered background light comes from the pump laser and could be blocked efficiently with a bandpass filter at the detector. However, the improvement is reduced by the fact that frequency conversion at ~50% efficiency and reduction in the number of photons will reduce the expected signal. A proper assessment of the advantages of frequency conversion would require a detailed assessment of the actual geometry and detector system proposed.

**Additional Considerations**. Further improvement of signal to noise can be obtained if the aperture of the detector system is increased to the order of 3 deg or 7.07 $deg^2$. Both the signal and the noise increase by a factor of ~ 7. However, this improves the counting statistics and the relative standard deviations decrease for the same number of shots. Clearly, considerable work has to be done in the proper design of the detection system, beam dumps and optical baffling of the interaction chamber for a successful experiment.

The contributions from other issues such as shot-to-shot timing jitter and pointing error also should be assessed in terms of their potential contribution to measurement error. However, it is expected that timing jitter within ± 30 fs could be achieved, which for the f/3 focal geometry of the pump would lead to a reduction in time-integrated interaction intensity squared of the pump up to ~10%. This would lead to lower measured values of the birefringence coefficient and would need to be corrected in the data analysis. It is expected that the shot-to-shot pointing error can be maintained within the order of 50% of the focal spot beam waist. The major error here comes from the varying intensity of the probe beam area overlapping with the pump due to lateral motion of the probe beam. The pointing error will lead to shot-to-shot fluctuations of scattered intensity down to ~60% of perfectly centered beams. The fractional scattering induced by the pump remains the same for any beam overlap within the central region of the probe. The effects of both shot-to-shot timing jitter and pointing error can be monitored throughout the series of shots using equivalent focal plane spatial and timing monitors. Correction factors can then be introduced in the data analysis to account for the average interaction intensity throughout the series of shots. The residual error for the average interaction intensities after applying the correction factors should be a fraction of the above deviations and are expected to be on the order of 10% or less after correction. This is of the same order of error that would exist in the calibration of the pump beam intensity itself.

Overall, it looks promising that a noise level of 0.1 ph/shot in a 1 $deg^2$ detector aperture can be achieved and thus even signals of 0.01 ph/shot/$deg^2$ could be measured by statistical accumulation of many shots. This would correspond to the case of using a pump laser at an intensity of the order of $10^{23}$ W/$cm^2$ outlined above. Assuming a silicon photodetector with a quantum efficiency of 80% at 800 nm is used and the overall optical efficiency of the combined confocal imaging and time gated detector system is 60% an overall quantum efficiency of detection of 48% per qvac scattered photon would be obtained. Thus, in $10^5$ shots (5000 detected noise photons and 500 detected qvac photons) an RMS error in the measured birefringence coefficient of 20% can be achieved. Such a measurement would be possible using a 1-Hz laser system over an operating period of 28 hours for data shots and 28 hours for time-shifted background noise shots, which could be carried out in a two-week period of operation at 5.6 hours of laser shots each day over 5 operational days. If the detector aperture is increased to 3 deg aperture (7.07 $deg^2$) then a statistical accuracy of ~8% can be achieved in the same counting period with $10^5$ shots. In this case, one could also consider using a streak camera for optical gating and the factor of 5.3 times reduction in detection quantum efficiency of the photocathode relative to a silicon detector would be compensated by the 9 times higher count rate. Such a streak camera system would be easier to implement than an optical Kerr shutter. Other errors such as timing jitter, spatial pointing error and the accuracy of intensity measurements themselves will then need to be taken into account and added into the overall measurement error.

An alternative way of looking at the result is the degree of confidence one would have in the actual verification of photon-photon scattering. Given an estimated background noise of 5000 counts the expected standard deviation in these counts would be $5000^{1/2}$ = 70.7 counts. The expected signal of 500 counts would then be 7 times this standard deviation in background noise giving a confidence level of 7σ for the actual detection of photon-photon scattering in this ideal case. In order to minimize any systematic errors due to drifts in alignment and laser pulse characteristics the null background shots (mistimed by 1 ps) should be interleaved with the real signal shots with short periods of acquisition for each. Note that the background scattered light primarily comes from linear scattering processes which are not very sensitive to the detailed wavefront or focal spot profile of the laser and thus is not expected to fluctuate significantly due to slight variations of laser conditions from shot to shot. However, it is recommended that the key laser conditions be monitored on a shot-by-shot basis, including, pulse shapes, timing, equivalent focal spot profiles and directional pointing of both beams, to determine the actual interaction conditions on each shot. This would allow rejection of shots which are poor and allow a weighted averaging of interaction intensities and conditions. The remaining variations in actual conditions from shot to shot are expected to add some additional noise to the measurement. Together with other deviations from the ideal conditions assumed for suppression of background scattered photons, we expect in a real experiment there would be an increase in the background noise. However, given the fact that we are starting from an estimated ideal signal to noise of $7\sigma$ a considerable increase in background noise can be tolerated while still allowing for a reasonable measurement to be achieved.

The viability of making a measurement can be improved either by reducing the noise level to less than 0.01 ph/shot or through using higher power lasers. If the noise is reduced by an order of magnitude, then the RMS error from a 28-hour signal period and 28-hour background period would be reduced to 7.7% for a 1 deg$^2$ detector and 2.9% for a 7.70 deg$^2$ detector (3 deg angular aperture). If higher laser intensities of $10^{24}$ W/cm$^2$ were available, the signal to noise would improve by an order of magnitude because the noise scales with intensity but the signal scales as the intensity squared. Thus, the counting times could be reduced by an order of magnitude, while still obtaining similar accuracies

## 9.- Conclusions.

We have presented a direct way to measure the QED nonlinear Lagrangian terms based on the collision of two counterpropagating optical lasers of ultra-high intensity. In spite of the extremely low value of the photon-photon cross section in the optical region, planned multi-PW lasers can make such an experiment feasible in the coming years. Other investigators have developed sophisticated tools to study a very general case (see for example Blinne et al. [57]). However, we have proposed here a specific paraxial or quasi-paraxial geometry and analyzed separately the contributions of the parallel and perpendicular polarizations, thus giving a direct insight on the coupling coefficients of the qvac Lagrangian for low-energy (optical) photons. In addition, we have analyzed the expected signal to noise in such an experiment indicating that such a measurement would be possible. To our knowledge this geometry is the most convenient one for a feasible nearterm future experiment using modern multipetawatt lasers currently in operation or under construction.

The challenge is to measure a few scattered photons due to the vacuum nonlinearities and separate this signal from the intrinsic noise of such extreme lasers. We conclude that with modern techniques to reduce noise level, time gating, ultrahigh vacuum and photon-counting detectors, such an experiment looks promising at pump intensities of $10^{24}$ W/cm$^2$ and barely feasible at today's intensity record of $10^{23}$ W/cm$^2$. The relevance of such measurement would be extraordinary because a new window may open to detect the existence of particles beyond The Standard Model of Particle Physics with a mass or charge smaller than the electron.

## Acknowledgments

The CLPU team acknowledges support from Laserlab Europe V Grant No. 871124, and from Junta de Castilla y León, Grant No. CLP087U16. The University of Maryland team acknowledges support from the National Science Foundation Grant # PHY2010392, SR acknowledges support from the Kulkarni Graduate Student Summer Research Fellowship. The University of Alberta team acknowledges support from the Natural Sciences and Engineering Research Council of Canada (RGPIN-2019-05013).